\documentclass[epsf,12pt]{article}

\usepackage[dvips]{graphicx}
\usepackage{epsfig}
\usepackage{psfig}
\usepackage{graphicx}
\usepackage{latexsym}
\usepackage[centertags]{amsmath}
\usepackage{amsfonts}
\usepackage{amssymb}
\usepackage{amsthm}
\usepackage{newlfont}

\begin{document}

\title
{ Four-loop verification of algorithm for Feynman diagrams
summation in $N=1$ supersymmetric electrodynamics.}

\author{A.B.Pimenov, K.V.Stepanyantz}

\maketitle

\begin{center}
{\em Moscow State University, physical faculty,\\
department of theoretical physics.\\
$119992$, Moscow, Russia}
\end{center}

\begin{abstract}
A method of Feynman diagrams summation, based on using
Schwinger-Dyson equations and Ward identities, is verified by
calculating some four-loop diagrams in $N=1$ supersymmetric
electrodynamics, regularized by higher derivatives. In particular,
for the considered diagrams correctness of an additional identity
for Green functions, which is not reduced to the gauge Ward
identity, is proved.
\end{abstract}

\sloppy


\section{Introduction.}
\hspace{\parindent}

Current indirect proofs of existence of the supersymmetry in the
Standard model make the problem of calculation of quantum
corrections in supersymmetric theories especially urgent. Due to
the supersymmetry the ultraviolet behavior of field theory models
is essentially improved. For example even in theories with
unextended supersymmetry it is possible to suggest the form of the
$\beta$-function exactly to all orders of the perturbation theory.
In the case of $N=1$ supersymmetric electrodynamics, which will be
considered in this paper, this $\beta$-function (that is called
the exact Novikov, Shifman, Vainshtein and Zakharov (NSVZ)
$\beta$-function) is \cite{NSVZ_Instanton}:

\begin{equation}\label{NSVZ_Beta}
\beta(\alpha) = \frac{\alpha^2}{\pi}\Big(1-\gamma(\alpha)\Big),
\end{equation}

\noindent where $\gamma(\alpha)$ is the anomalous dimension of the
matter superfield.

Explicit calculations, made with the dimensional reduction
\cite{Siegel}, confirm this proposal, but require a special choice
of the subtraction scheme \cite{Tarasov,North}. Explicit
calculations in two- \cite{hep,tmf2} and three-loop
\cite{ThreeLoop} approximations for the $N=1$ supersymmetric
electrodynamics with the higher derivative regularization
\cite{Slavnov,Bakeyev} reveal that renormalization of the operator
$W_a C^{ab} W_b$ is exhausted at the one-loop and the
Gell-Mann-Low function coincides with the exact NSVZ
$\beta$-function and has corrections in all orders of the
perturbation theory.

An attempt to perform explicit calculations exactly to all orders
of the perturbation theory was made in Ref. \cite{SD}. According
to this paper it is possible to calculate a large number of
Feynman diagrams exactly to all orders of the perturbation theory
using Schwinger-Dyson equations and Ward identities. Nevertheless,
some diagrams can not be obtained by this method. However,
explicit calculations (up to the three-loop approximation
inclusive) show, that the sum of undefined contributions is 0.
This means existence of an identity, which is not reduced to the
Ward identities and can be graphically written in the form
(\ref{Graphical_Identity}). (In the analitic form this identity is
given by Eq. (\ref{New_Identity})). It relates the two-point Green
function of a matter superfield and a three-point Green function,
in which external lines correspond to a gauge field, a chiral
matter superfield and an external source introduced in a special
way. This source is an arbitrary scalar superfield.

Identity (\ref{New_Identity}) is important because it restricts
nontrivially the structure of the divergences of the theory.
Really, due to the supersymmetry and gauge invariance the
two-point Green function of the gauge field should be proportional
to $V\partial^2\Pi_{1/2} V$, the coefficient being an unknown
function. (It is a supersymmetric analog of the transversality
condition in the ordinary electrodynamics.) Using the
Schwinger-Dyson equations and Ward identities it is possible to
rewrite this function via the two-point Green function of the
matter superfields and another function, which is not fixed by the
gauge invariance. Actually this means, that the two-point Green
function of the gauge superfield is equivalently rewritten in a
different form, as earlier, up to an undefined (from the Ward
identities or, equivalently, from the gauge invariance) function.
However, according to new identity the terms, which depend on the
unknown function, are actually equal to zero. So this identity
really removes the arbitrariness in the structure of the
divergence.

Certainly, it is highly desirable to find a true reason of this
identity -- for example, some symmetry. But first it is necessary
to verify, that it really takes place and is not caused by an
accidental coincidence.

The considered identity is nontrivial starting from the three-loop
approximation. In Ref. \cite{Identity} it was proved for a special
class of diagrams exactly to all orders of the perturbation
theory. Such diagrams contain the only loop of the matter
superfields and any two cuts of this loop do not make the diagram
disconnected. (In this case the technical part of the proof is
simpler.) Possibly in the general case the proof can be made
similarly. In order to confirm that such a proof is possible, it
is desirable to verify identity (\ref{New_Identity}) for other
diagrams. This requires making explicit four-loop calculations.
These calculations are made in this paper for a special group of
four-loop diagrams, which can be made disconnected by two cuts of
the matter superfields loop.

The paper is organized as follows:

In Sec. \ref{Section_SUSY_QED} the basic information about $N=1$
supersymmetric electrodynamics and its regularization by higher
derivatives is reminded. The method of summation Feynman diagrams,
based on using Schwinger-Dyson equations and Ward identities is
described in Sec. \ref{Section_SD}. In this section we write an
identity, to which the Green functions are supposed to satisfy.
Four-loop verification of the results, presented in Sec.
\ref{Section_SD}, is made in Sec. \ref{Section_Four_Loop}. A brief
summary of the paper is presented in the Conclusion.


\section{$N=1$ supersymmetric electrodynamics and higher
derivative regularization.} \label{Section_SUSY_QED}
\hspace{\parindent}

The massless $N=1$ supersymmetric electrodynamics with the higher
derivatives term in the superspace is described by the following
action:

\begin{equation}\label{Regularized_SQED_Action}
S = \frac{1}{4 e^2} \mbox{Re}\int d^4x\,d^2\theta\,W_a C^{ab}
\Big(1+ \frac{\partial^{2n}}{\Lambda^{2n}}\Big) W_b +
\frac{1}{4}\int d^4x\, d^4\theta\, \Big(\phi^* e^{2V}\phi
+\tilde\phi^* e^{-2V}\tilde\phi\Big).
\end{equation}

\noindent Here $\phi$ and $\tilde\phi$ are the chiral matter
superfields and $V$ is a real scalar superfield, which contains
the gauge field $A_\mu$ as a component. The superfield $W_a$ is a
supersymmetric analog of the stress tensor of the gauge field. In
the Abelian case it is defined by

\begin{equation}
W_a = \frac{1}{16} \bar D (1-\gamma_5) D\Big[(1+\gamma_5)D_a
V\Big],
\end{equation}

\noindent where $D$ is a supersymmetric covariant derivative. It
is important to note, that in the Abelian case the superfield
$W^a$ is gauge invariant, so that action
(\ref{Regularized_SQED_Action}) will be also gauge invariant.

Quantization of model (\ref{Regularized_SQED_Action}) can be made
by the standard way. For this purpose it is convenient to use the
supergraphs technique, described in book \cite{West} in details,
and to fix the gauge invariance by adding the following terms:

\begin{equation}\label{Gauge_Fixing}
S_{gf} = - \frac{1}{64 e^2}\int d^4x\,d^4\theta\, \Bigg(V D^2 \bar
D^2 \Big(1 + \frac{\partial^{2n}}{\Lambda^{2n}}\Big) V + V \bar
D^2 D^2 \Big(1+ \frac{\partial^{2n}}{\Lambda^{2n}}\Big) V\Bigg),
\end{equation}

\noindent where

\begin{equation}
D^2 \equiv \frac{1}{2} \bar D (1+\gamma_5)D;\qquad \bar D^2 \equiv
\frac{1}{2}\bar D (1-\gamma_5) D.
\end{equation}

\noindent After adding such terms a part of the action, quadratic
in the superfield $V$, will have the simplest form

\begin{equation}
S_{gauge} + S_{gf} = \frac{1}{4 e^2}\int d^4x\,d^4\theta\,
V\partial^2 \Big(1+ \frac{\partial^{2n}}{\Lambda^{2n}}\Big) V.
\end{equation}

\noindent In the Abelian case, considered here, diagrams
containing ghost loops are absent.

It is well known, that adding of the higher derivative term does
not remove divergences in one-loop diagrams. In order to
regularize them, it is necessary to insert in the generating
functional the Pauli-Villars determinants \cite{Slavnov_Book}.

The generating functional can be written in the form

\begin{equation}\label{Modified_Z}
Z = \int DV\,D\phi\,D\tilde \phi\, \prod\limits_i \Big(\det
PV(V,M_i)\Big)^{c_i}
\exp\Big(i(S_{ren}+S_{gf}+S_S+S_{\phi_0})\Big).
\end{equation}

\noindent Here

\begin{eqnarray}\label{Renormalized_Action}
&& S_{ren} = \frac{1}{4 e^2} Z_3(e,\Lambda/\mu)\, \mbox{Re}\int
d^4x\,d^2\theta\,W_a C^{ab} \Big(1+
\frac{\partial^{2n}}{\Lambda^{2n}}\Big) W_b
+\nonumber\\
&& \qquad\qquad\qquad\qquad\qquad +
Z(e,\Lambda/\mu)\,\frac{1}{4}\int d^4x\, d^4\theta\, \Big(\phi^*
e^{2V}\phi +\tilde\phi^* e^{-2V}\tilde\phi\Big)\qquad
\end{eqnarray}

\noindent is the renormalized action, the action for the gauge
fixing terms are given by Eq. (\ref{Gauge_Fixing}) (It is
convenient to substitute $e$ by the  bare charge $e_0$ in it), the
Pauli-Villars determinants are defined by

\begin{equation}\label{PV_Determinants}
\Big(\det PV(V,M)\Big)^{-1} = \int D\Phi\,D\tilde \Phi\,
\exp\Big(i S_{PV}\Big),
\end{equation}

\noindent where

\begin{eqnarray}
&& S_{PV}\equiv Z(e,\Lambda/\mu) \frac{1}{4} \int
d^4x\,d^4\theta\, \Big(\Phi^* e^{2V}\Phi + \tilde\Phi^*
e^{-2V}\tilde\Phi \Big)
+\qquad\nonumber\\
&&\qquad\qquad\qquad\qquad\qquad  + \frac{1}{2}\int
d^4x\,d^2\theta\, M \tilde\Phi \Phi + \frac{1}{2}\int
d^4x\,d^2\bar\theta\, M \tilde\Phi^* \Phi^*,\qquad
\end{eqnarray}

\noindent and the coefficients $c_i$ satisfy conditions

\begin{equation}
\sum\limits_i c_i = 1;\qquad \sum\limits_i c_i M_i^2 = 0.
\end{equation}

\noindent Below we will assume, that $M_i = a_i\Lambda$, where
$a_i$ are some constants. Insertion of Pauli-Villars determinants
allows to cancel remaining divergences in all one-loop diagrams,
including diagrams, containing insertions of counterterms.

The terms with sources are written in the form

\begin{eqnarray}\label{Sources}
&& S_S = \int d^4x\,d^4\theta\,J V + \int d^4x\,d^2\theta\,
\Big(j\,\phi + \tilde j\,\tilde\phi \Big) + \int
d^4x\,d^2\bar\theta\, \Big(j^*\phi^* + \tilde j^*
\tilde\phi^*\Big).
\end{eqnarray}

\noindent Moreover, in generating functional (\ref{Modified_Z}) we
introduced the expression

\begin{equation}
S_{\phi_0} = \frac{1}{4}\int d^4x\,d^4\theta\,\Big(\phi_0^*\,
e^{2V} \phi + \phi^*\, e^{2V} \phi_0 + \tilde\phi_0^*\,
e^{-2V}\tilde\phi + \tilde\phi^*\, e^{-2V}\tilde\phi_0 \Big),
\end{equation}

\noindent where $\phi_0$, $\phi_0^*$, $\tilde\phi_0$ and
$\tilde\phi_0^*$ are scalar superfields. They are some parameters,
which are not chiral or antichiral. In principle, it is not
necessary to introduce the term $S_{\phi_0}$ into the generating
functional, but the presence of the parameters $\phi_0$ is highly
desirable for the investigation of Schwinger-Dyson equations.

In our notations the generating functional for the connected Green
functions is written as

\begin{equation}\label{W}
W = - i\ln Z,
\end{equation}

\noindent and an effective action is obtained by making a Legendre
transformation:

\begin{equation}\label{Gamma}
\Gamma = W - \int d^4x\,d^4\theta\,J V - \int d^4x\,d^2\theta\,
\Big(j\,\phi + \tilde j\,\tilde\phi \Big) - \int
d^4x\,d^2\bar\theta\, \Big(j^*\phi^* + \tilde j^* \tilde\phi^*
\Big),
\end{equation}

\noindent where the sources $J$, $j$ and $\tilde j$ is to be
eliminated in terms of the fields $V$, $\phi$ and $\tilde\phi$,
through solving equations

\begin{equation}
V = \frac{\delta W}{\delta J};\qquad \phi = \frac{\delta W}{\delta
j};\qquad \tilde\phi = \frac{\delta W}{\delta\tilde j}.
\end{equation}


\section{Summation of Feynman diagrams in $N=1$ supersymmetric quantum
electrodynamics} \label{Section_SD} \hspace{\parindent}

From generating functional (\ref{Modified_Z}) it is possible to
obtain \cite{SD} the Schwinger-Dyson equations, which are
graphically written as

\begin{equation}\label{SD_Equation}
\begin{picture}(0,1.8)
\put(-5,0.6){$\Gamma^{(2)}_V =$} \put(0.8,0.6){+} \hspace*{-3.5cm}
\includegraphics[scale=0.88]{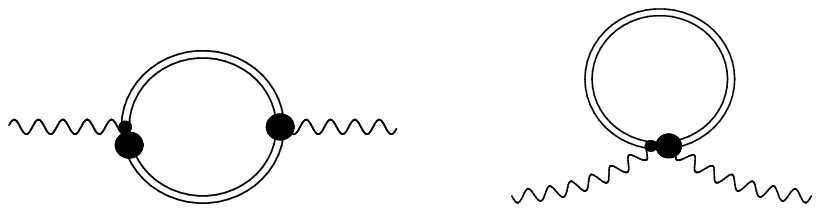}
\end{picture}
\end{equation}

\noindent where $\Gamma^{(2)}_V$ is a two-point Green function of
the gauge field. We will always set the renormalization constant
$Z=1$, because the dependence on $Z$ can be easily restored in the
final result and this dependence is not essential in this paper.

The double line in Eq. (\ref{SD_Equation}) denotes the exact
propagator ($Z=1$)

\begin{equation}\label{Inverse_Functions}
\Bigg(\frac{\delta^2\Gamma}{\delta\phi_x^*\delta\phi_y}\Bigg)^{-1}
= -  \frac{D_x^2 \bar D_x^2}{4 \partial^2 G} \delta^8_{xy},
\end{equation}

\noindent in which the function $G(q^2)$ is defined by the
two-point Green function as follows:

\begin{equation}\label{Explicit_Green_Functions}
\frac{\delta^2\Gamma}{\delta\phi_x^*\delta\phi_y} = \frac{D_x^2
\bar D_x^2}{16} G(\partial^2) \delta^8_{xy},
\end{equation}

\noindent where $\delta^8_{xy}\equiv
\delta^4(x-y)\delta^4(\theta_x-\theta_y)$, and the lower indexes
denote points, in which considered expressions are taken.

The large circle denotes the effective vertex, which is written as
\cite{SD}

\begin{eqnarray}\label{Vertex2}
&& \frac{\delta^3\Gamma}{\delta
V_x\delta\phi_y\delta\phi^*_{z}}\Bigg|_{p=0} =
\partial^2\Pi_{1/2}{}_x\Big(\bar D_x^2\delta^8_{xy} D_x^2
\delta^8_{xz}\Big) F(q^2) +\nonumber\\
&& \qquad\qquad\quad +\frac{1}{32} q^\mu G'(q^2) \bar
D\gamma^\mu\gamma_5 D_x \Big(\bar D_x^2\delta^8_{xy} D_x^2
\delta^8_{xz}\Big) + \frac{1}{8} \bar D_x^2\delta^8_{xy} D_x^2
\delta^8_{xz}\, G(q^2),
\end{eqnarray}

\noindent due to the Ward identities. The strokes denote
derivatives with respect to $q^2$,

\begin{equation}
\Pi_{1/2} = - \frac{1}{16 \partial^2} D^a \bar D^2 D_a = -
\frac{1}{16 \partial^2} \bar D^a D^2 \bar D_a
\end{equation}

\noindent is a supersymmetric transversal projector, and $F(q^2)$
is a function, which can not be defined from the Ward identities.
Here

\begin{eqnarray}
&& D^a \equiv \Big[\frac{1}{2}\bar D (1+\gamma_5)\Big]^a;\qquad
D_a \equiv \Big[\frac{1}{2}(1+\gamma_5) D\Big]_a;\nonumber\\
&& \bar D^a \equiv \Big[\frac{1}{2}\bar D (1 -
\gamma_5)\Big]^a;\qquad \bar D_a \equiv
\Big[\frac{1}{2}(1-\gamma_5) D\Big]_a.
\end{eqnarray}

Two adjacent circles denote an effective vertex, consisting of 1PI
diagrams, in which one of the external lines is attached to the
very left edge. Such vertexes are given by \cite{SD}

\begin{equation}\label{Useful_Identities}
\frac{\delta^2\Gamma}{\delta\phi_y \delta \phi_{0z}^*} =
\frac{1}{4} \frac{\delta}{\delta \phi_y}
\exp\Bigg(\frac{2}{i}\frac{\delta}{\delta J_z} + 2 V_z\Bigg)
\phi_z = - \frac{1}{8} G \bar D_y^2\delta^8_{yz}
\end{equation}

\noindent in the case of one external $V$-line (the vertex in the
first diagram of Eq. (\ref{SD_Equation})) and

\begin{eqnarray}\label{Vertex3}
&& \frac{\delta^3\Gamma}{\delta
V_x\delta\phi_y\delta\phi^*_{0z}}\Bigg|_{p=0} = \frac{1}{4}
\frac{\delta}{\delta V_x}\frac{\delta}{\delta \phi_y}
\exp\Bigg(\frac{2}{i}\frac{\delta}{\delta J_z}+2V_z\Bigg)
\phi_z\Bigg|_{p=0} =\nonumber\\
&& = -2 \partial^2\Pi_{1/2}{}_x\Big(\bar D_x^2\delta^8_{xy}
\delta^8_{xz}\Big) F(q^2) + \frac{1}{8} D^a C_{ab} \bar
D_x^2\Big(\bar D_x^2\delta^8_{xy} D_x^b \delta^8_{xz} \Big) f(q^2)
+\vphantom{\frac{1}{2}}\nonumber\\
&&\qquad\qquad\qquad\qquad -\frac{1}{16} q^\mu G'(q^2) \bar
D\gamma^\mu\gamma_5 D_x \Big(\bar D_x^2\delta^8_{xy}
\delta^8_{xz}\Big) -\frac{1}{4} \bar D_x^2\delta^8_{xy}
\delta^8_{xz}\, G(q^2),
\end{eqnarray}

\noindent in the case of two external $V$-lines (in the second
diagram of Eq. (\ref{SD_Equation})). Here $f(q^2)$ is one more
function, which can not be found from the Ward identity.

The expression

\begin{equation}\label{Limit}
\frac{d}{d\ln\Lambda}\Gamma^{(2)}_V\Bigg|_{p=0},
\end{equation}

\noindent can be calculated \cite{SD} by substituting solutions of
Ward identities into the Schwinger-Dyson equations.

A result of calculation of any diagram can be written in the form

\begin{equation}
\int \frac{d^4p}{(2\pi)^4}\, d^4\theta\,\Big(V\partial^2\Pi_{1/2}
V A(p^2) + V^2 B(p^2)\Big),
\end{equation}

\noindent where $A$ and $B$ are some functions. It is actually
(with a small modification) a supersymmetric analog of expansion
to the longitudinal and transversal parts. Due to the Ward
identity the sum of diagrams contains only transversal parts, i.e.
terms, proportional to $V\partial^2 \Pi_{1/2} V$. It is confirmed
by calculations, made in Ref. \cite{SD}: Terms, proportional to
$V^2$, are 0 in the sum of diagrams, and we will not write them
here. Terms, proportional to $V\partial^2 \Pi_{1/2} V$, in the
effective diagrams are given by

\begin{equation}\label{First_Diagram}
\begin{picture}(0,1)
\put(-7,-0.5){\includegraphics[scale=0.4]{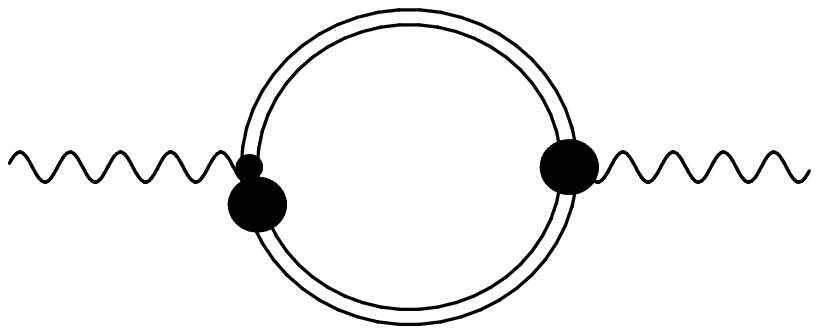}}\hspace{-3.6cm}
${\displaystyle = V\partial^2\Pi_{1/2} V \frac{d}{d\ln\Lambda}
\int \frac{d^4q}{(2\pi)^4}\Bigg(\frac{8 F}{q^2 G}  +
\frac{1}{2q^2}\,\frac{d}{dq^2}\ln\Big(q^2 G^2\Big)\Bigg);}$
\end{picture}
\end{equation}

\begin{equation}\label{Second_Diagram}
\begin{picture}(0,1)
\put(-6.7,-0.8){\includegraphics[scale=0.4]{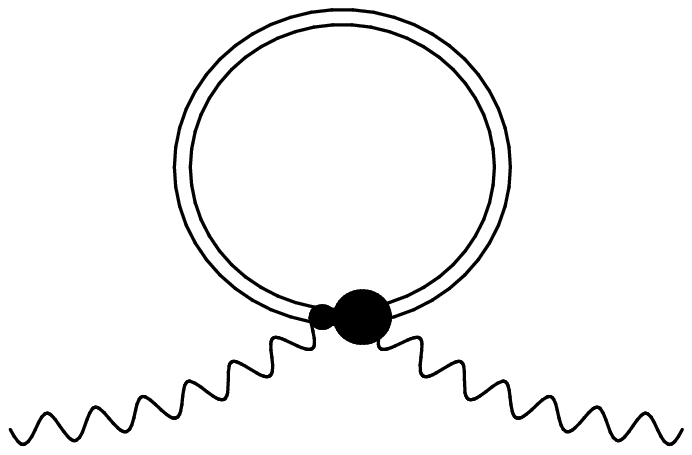}}\hspace{-3.6cm}
${\displaystyle = V\partial^2\Pi_{1/2} V \frac{d}{d\ln\Lambda}
\int \frac{d^4q}{(2\pi)^4}\Bigg(-\frac{8 F}{q^2 G} - \frac{8
f}{q^2 G} \Bigg).}$
\end{picture}
\end{equation}

\bigskip

\noindent (Here we omit integration with respect to the external
momentum and external $\theta$.) Expressions for these diagrams
also contain terms, proportional to $V^2$ and similar terms with
the Pauli-Villars fields, which we omitted for brevity. Their form
can be found in Ref. \cite{SD}. Terms with the Pauli-Villars
fields give additional contributions, which cancel the remaining
divergences. Investigation of them is a bit more complicated and
requires making similar calculations in the massive theory. We
will always assume existence of terms with the Pauli-Villars
fields, but will not write them explicitly.

Note now, that terms, containing the unknown function $F$, are
completely cancelled in the sum of contributions
(\ref{First_Diagram}) and (\ref{Second_Diagram}). However there
are terms, containing the unknown function $f$:

\begin{equation}\label{Result}
\frac{d}{d\ln\Lambda}\,\frac{\delta^2\Delta\Gamma}{\delta
V_x\,\delta V_y} \Bigg|_{p=0} = 2 \partial^2\Pi_{1/2}\delta^8_{xy}
\int\frac{d^4q}{(2\pi)^4}\,\Bigg(\frac{1}{2q^2} \frac{d}{dq^2} \ln
(q^2 G^2) - \frac{8 f}{q^2 G} - (PV)\Bigg),
\end{equation}

\noindent where $(PV)$ denotes similar terms with the
Pauli-Villars fields. Nevertheless, explicit three-loop
calculations show that the following identity takes place:

\begin{equation}\label{New_Identity}
\frac{d}{d\ln\Lambda} \int \frac{d^4q}{(2\pi)^4} \frac{f(q^2)}{q^2
G(q^2)} = 0,
\end{equation}

\noindent where the function $G$ is defined from the two-point
Green function of the matter superfield by Eq.
(\ref{Explicit_Green_Functions}), and the function $f$ -- from the
three-point Green function with the external lines $V$, $\phi$ and
$\phi_0^*$ by Eq. (\ref{Vertex3}).

An attempt to prove identity (\ref{New_Identity}) exactly to all
orders was made in Ref. \cite{Identity} by approximately the same
method, which is used for the proof of the Ward identity diagram
by diagram. Although application of this method appears to be
possible, the proof was not made in the general case due to the
technical difficulties. It was made only for diagrams, which
contain the only loop of the matter superfields and remain
connected after any two cuts of this loop. In the general case the
proof, made by the same method, seems to be possible, but much
more complicated technically.

Identity (\ref{New_Identity}) can be graphically written as

\bigskip

\begin{equation}\label{Graphical_Identity}
\begin{picture}(0,2)
\put(0.5,2){$(I_1)^a$} \put(-0.3,-0.4){$D_a V_x$}
\put(2.8,-0.4){$V_y$}
\end{picture}
\includegraphics[scale=0.4]{effdg6.eps}
\put(0,0.9){= 0,}
\end{equation}

\bigskip

\noindent that was also proved in Ref. \cite{Identity}. The symbol
$(I_1)^a$ (here we use notations of Ref. \cite{Identity}) means,
that the double line in this case corresponds to the expression

\begin{equation}
- \frac{1}{2 \partial^2 G(\partial^2)}D^a_x \bar D^2_x
\delta^8_{xz},
\end{equation}

\noindent instead of the effective propagator
(\ref{Inverse_Functions}), which is proportional to $D^2_x \bar
D^2_x \delta^8_{xz}$. Note, that equality (\ref{New_Identity}) is
not valid in the massive case. Its corresponding modification can
be found in \cite{SD}.

\section{Four-loop calculations} \label{Section_Four_Loop}
\hspace{\parindent}

\begin{figure}[h]
    \centering
    \includegraphics[width=0.6in]{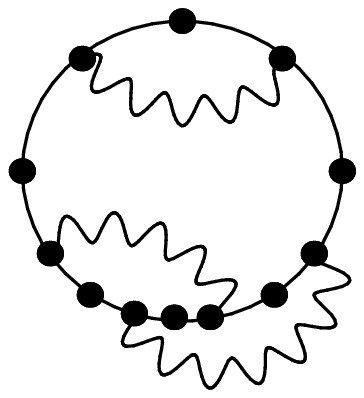}
    \caption{Schematic picture of the class of diagrams, considered
    in the paper}
    \label{explain}
\end{figure}

Note now, that expressions (\ref{First_Diagram}) and
(\ref{Second_Diagram}) allow to find not only sums of all diagrams
with two external lines of the gauge field, but also sums of
special classes of such diagrams. Such classes of diagrams are
obtained from a frame, to which external lines are attached by all
possible ways. For example, in Fig. \ref{explain} we present a
frame for diagrams, which are investigated in this paper. The dots
denote all possible points, to which two external lines of the
gauge field can be attached.

If a frame contains the only loop of the matter superfield and can
not be made disconnected by two cuts of this loop, then for the
corresponding class of diagrams identity
(\ref{Graphical_Identity}) was proved in Ref. \cite{Identity}.
However it is desirable to verify, if this identity takes place in
the other cases. For this purpose we consider diagrams, which are
obtained from the frame, presented in Fig. \ref{explain}, by
adding two external lines. (Such diagrams can be evidently made
disconnected by two cuts of the loop of matter superfields.)

We will calculate the considered diagrams by two ways:

1. using Eqs. (\ref{First_Diagram}) and (\ref{Second_Diagram}) and
the functions $G$, $f$ and $F$, obtained preliminary.

2. by explicit calculation using the supergraph method.

Thus it is possible to perform a four-loop verification of Eqs.
(\ref{First_Diagram}) and (\ref{Second_Diagram}), and also
identity (\ref{New_Identity}).

In order to find diagrams which will be essential for finding the
unknown functions $G$, $f$ and $F$ in the considered case, it is
convenient to use the following simple speculations: A diagram,
presented in Fig. \ref{explain}, can be considered as a formal
product of one- and two-loop diagrams with $\phi^*$ and $\phi$
external lines (pairs of their ends are identified):

\begin{figure}[h]
\centering \vspace{0.4cm}
\includegraphics[width=0.6in]{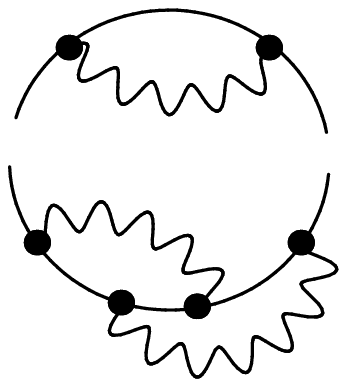}
\end{figure}

\noindent or as a three-loop diagram with the identified ends:

\begin{figure}[h]
\centering \vspace{0.6cm}
\includegraphics[width=0.6in]{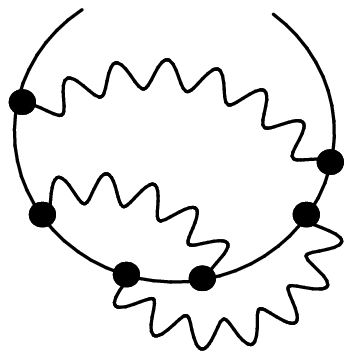} \hspace{2cm}
\includegraphics[width=0.6in]{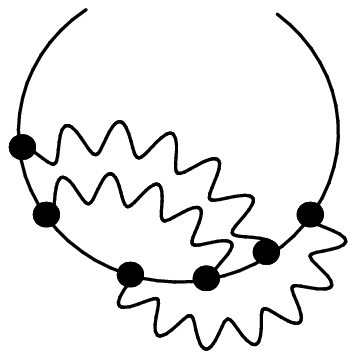}
\end{figure}

\noindent The parts of the diagram, obtained by this way, are the
Feynman diagrams for finding the function $G$. To define the
functions $f$ and $F$ it is necessary to add one more external
line of the superfield $V$ to such diagrams.

\begin{figure}[h]
    \centering
    \includegraphics{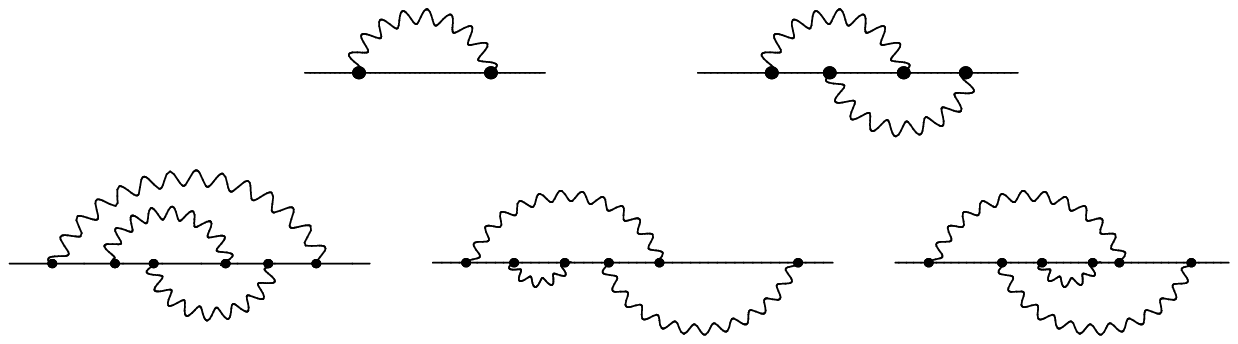}
    \caption{The diagrams, used for calculation of the function $G$}
    \label{1-2-3_loop_anomal}
\end{figure}

Thus it order to find the function $G$ in the considered case it
is necessary to calculate diagrams presented in Fig.
\ref{1-2-3_loop_anomal}. Then the function $G$ is obtained
according to definition (\ref{Explicit_Green_Functions}). One-,
two- and three-loop parts of the function $G$, defined by diagrams
in Fig. \ref{1-2-3_loop_anomal}, will be denoted by $G_1$, $G_2$
and $G_3$ respectively. Expressions for them are presented in
Appendix \ref{Appendix_Functions}. The complete function $G$
(certainly without diagrams, which are not essential for this
paper) in the considered approximation is given by

\begin{equation}\label{G}
G(q^2) = 1 + G_1 + G_2 + G_3,
\end{equation}

\noindent where the unity is a tree contribution.

The functions $F(q^2)$ and $f(q^2)$ are defined from three-loop
Green functions by Eq. (\ref{Vertex3}). As in the case of the
function $G$, in order to obtain the contribution, which
corresponds to the considered class of the four-loop diagrams, it
is sufficient to calculate only diagrams of a special form. These
diagrams are obtained from diagrams, presented in Fig.
\ref{1-2-3_loop_anomal}, by all possible insertions of one
external $V$-line.

We will denote one-, two- and three-loop contributions to the
function $f$  by $f_1$, $f_2$ and $f_3$ respectively. Similar
notation we will use for the function $F$. Because in the tree
approximation these functions are 0 and $f_1=0$, in the considered
order we have

\begin{equation}
F = F_1 + F_2 + F_3;\qquad f = f_2+ f_3.
\end{equation}

\noindent Expressions for $f_1$, $f_2$, $f_3$, $F_1$ and $F_2$,
obtained by calculation of the above pointed diagrams, are
presented in Appendix \ref{Appendix_Functions}. An expression for
$F_3$ has been calculated, but it is not presented because it is
very large. (It is not required for verification of identity
(\ref{New_Identity}).)

Using the obtained expressions it is possible to verify identity
(\ref{New_Identity}). With the considered accuracy we have

\begin{equation}\label{-8fG}
\frac{f}{q^2 G} = \frac{1}{q^2 } \Big(f_3 - f_2\cdot G_1\Big).
\end{equation}

\noindent Substituting here expressions for the functions $G_1$,
$f_2$ and $f_3$ from Eqs. (\ref{G_1}), (\ref{f-2}), (\ref{f-3}) we
obtain, that the integrand can be written as a total derivative
with respect to the momentum $q$:

\begin{eqnarray}\label{f-total_deriv}
&& \int \frac{d^4q}{(2\pi)^4}\,\frac{f}{q^2 G} = i e^6 \int
\frac{d^4q\, d^4k\, d^4l\, d^4r }{(2\pi)^{16}} \left[ 1+(-1)^n
\frac{k^{2n}}{\Lambda^{2n}} \right]^{-1}\left[ 1+(-1)^n
\frac{l^{2n}}{\Lambda^{2n}}
\right]^{-1}\times\qquad\nonumber\\
&& \times \left[ 1+(-1)^n \frac{r^{2n}}{\Lambda^{2n}} \right]^{-1}
\frac{\partial}{\partial q^{\mu}}\left\{ \frac{(2q+k+l)^{\mu}}{k^2
l^2 r^2 q^2 (k+q)^2 (q+l)^2 (q+r)^2 (k+q+l)^2} \right\}.
\end{eqnarray}

\noindent This equality can be verified by calculating the
derivative with respect to $q^\mu$ using the Leibnitz rule and
comparing the result with Eq. (\ref{-8fG}), in which expressions
for the functions $G_1$, $f_2$ and $f_3$ are obtained by explicit
calculation of diagrams.

Because the integrand in Eq. (\ref{f-total_deriv}) is a total
derivative with respect to $q^{\mu}$ of the expression, which goes
to 0 in the limit $q\to\infty$, expression (\ref{f-total_deriv})
is 0. This means that identity (\ref{New_Identity}) is correct in
the considered approximation and for the considered class of
diagrams.

Note, that in this case using of the higher derivative
regularization is essential. In spite of the parameter $\Lambda$
is encountered only in propagators of the superfield $V$, which do
not contain the momentum $q$, this regularization allows to make
calculations in the limit $p\to 0$. It follows from existence of
limit (\ref{Limit}), which was proved in Ref. \cite{SD}. Taking
this limit is senseless without the higher derivative
regularization, because this produces infinities. Another
important reason of using the higher derivative regularization is
the possible problems with integration with respect to the
momentum $q$ in Eq. (\ref{Result}) in $D\ne 4$ dimensions. For
example, it is possible, that in $D$ dimensions expression
(\ref{-8fG}) can not be presented as a total derivative. In
principle, identity (\ref{New_Identity}) can be considered as a
formal relation in the non-regularized theory, which is made
sensible by the regularization. However, most likely it is not
valid for an arbitrary regularization in the regularized theory.
Because supersymmetric theories are usually regularized either by
the dimensional reduction or by the higher derivative
regularization, it is sensible to say, that identity
(\ref{New_Identity}) is valid in the theory, regularized by higher
derivatives.

It is also desirable to verify the method of summation of Feynman
diagrams by using the Schwinger-Dyson equations and Ward
identities. For this purpose it is possible to calculate both
effective diagrams in Eq. (\ref{SD_Equation}) in the four-loop
approximation explicitly and compare the result with Eqs.
(\ref{First_Diagram}) and (\ref{Second_Diagram}). In considered
approximation

\begin{eqnarray}
&& \frac{1}{q^2}\,\frac{d}{dq^2}\ln(G^2) = \frac{q^{\mu}}{q^4}\,
\frac{\partial}{\partial q^{\mu}}\Big( G_3
-G_1 \cdot G_2 \Big);\\
&& \frac{F}{q^2 G} = \frac{1}{q^2 } \Big(F_3 - F_2\cdot G_1-
F_1\cdot G_2\Big).
\end{eqnarray}

\noindent Explicit calculations were made by the method, proposed
in Ref. \cite{Vestnik}, which simplifies finding a part of
diagram, proportional to $V\partial^2 \Pi_{1/2}V$. Nevertheless,
it does not allow to obtain a part, proportional to $V^2$. That is
why the verification of Eqs. (\ref{First_Diagram}) and
(\ref{Second_Diagram}) was made only for a part, proportional to
$V\partial^2\Pi_{1/2} V$. In the both cases this verification
completely confirms them.

\begin{figure}[h]
    \centering
    \includegraphics[width=1.5in]{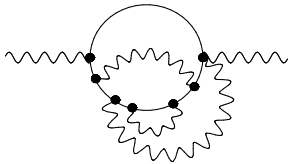}\\
    \caption{One of the diagrams, giving contribution to both
    effective diagrams}\label{Figure_Example}
\end{figure}

As a small technical remark let us note, that making this
verification it was necessary to take into account, that a large
number of ordinary Feynman diagrams contributed both to the first
effective diagram in Eq. (\ref{SD_Equation}) and to the second
one. For example, it is easy to see, that a contribution of
diagram, presented in Fig. \ref{Figure_Example}

\noindent is divided into parts $3/4$ and $1/4$, which correspond
to the first and to the second diagrams in Eq.
(\ref{SD_Equation}).


\section{Conclusion}
\label{Section_Conclusion} \hspace{\parindent}

In this paper the four-loop verification of identity
(\ref{New_Identity}) has been made for a special class of Feynman
diagrams. The proof, presented in Ref. \cite{Identity}, is not
applicable for these diagrams, because structure of the diagrams
sum is more complicated and there are additional restrictions to
possible positions of the gauge field external lines. Due to the
confirmation of the considered identity, obtained here, this
identity seems to be also valid for arbitrary diagrams. Its origin
seems to be a symmetry of the theory. However, this symmetry
surely is not the gauge invariance.

To conclude we stress again, that the especial importance of the
obtained identity is that it essentially restricts the form of
divergences in the considered theory. The matter is that it is
possible to reduce the expression for the two-point Green function
of the gauge field to the expression, depending only on the
two-point Green function of the matter superfield and LHS of this
identity, using only the gauge invariance. Thus the considered
identity removes all undefined expressions from the two-point
Green function of the gauge field and relates it with the
two-point Green function for the matter superfield.

\bigskip
\bigskip

\noindent {\Large\bf Acknowledgements.}

\bigskip

\noindent This work was partially supported by Russian Foundation
for Basic Research (grant No 05-01-00541).

\appendix

\section{Explicit expressions for the functions $G$, $f$ and $F$}
\label{Appendix_Functions} \hspace{\parindent}

Below we present expressions for the functions $G$, $f$ and $F$.
The lower index shows in what order of the perturbation theory
(for diagrams of the considered class) the corresponding function
was calculated.

\begin{eqnarray}\label{G_1}
&& G_1 = 2\, i e^2 \int \frac{d^4 k}{(2\pi)^4} \,\left[ 1+(-1)^n
\frac{k^{2n}}{\Lambda^{2n}} \right]^{-1} \, \frac{1}{k^2\,
(q+k)^2}; \\
\label{G_2} && G_2 = 4\,e^4 \int
\frac{d^4k\,d^4l}{(2\pi)^8}\,\left[ 1+(-1)^n
\frac{k^{2n}}{\Lambda^{2n}} \right]^{-1} \left[ 1+(-1)^n
\frac{l^{2n}}{\Lambda^{2n}}
\right]^{-1} \vphantom{\Bigg(}\times \\
&&\times  \frac{(2q + k +l)^2}{k^2\, l^2 (k+q)^2\, (q+l)^2\,
(k+q+l)^2};\nonumber\\
\label{G_3} && G_3 = -4\, i e^6 \int
\frac{d^4k\,d^4l\,d^4r}{(2\pi)^{12}}\,\left[1+(-1)^n
\frac{k^{2n}}{\Lambda^{2n}} \right]^{-1} \left[ 1+(-1)^n
\frac{l^{2n}}{\Lambda^{2n}} \right]^{-1} \times\nonumber\\
&&\times \left[ 1+(-1)^n \frac{r^{2n}}{\Lambda^{2n}} \right]^{-1}
\frac{1}{k^2 l^2 r^2 (q+k)^2 (q+l)^2 (q+k+l)^2}
\times \nonumber \\
&& \times\Bigg(\frac{(2q + 2k + r +l)^2 (q+l)^2}{
(q+r+k)^2\,(q+r+k+l)^2} + \frac{(2q + k +l)^2}{ (q+r+k+l)^2} +
\frac{2 (2q + k +l)^2}{(q+r+k)^2} \Bigg).\qquad
\end{eqnarray}

\begin{eqnarray}\label{f-1}
&& f_1 (q^2) = 0;\vphantom{\Big(}\\
\label{f-2} && f_2(q^2) = \frac{1}{2}\, e^4\int \frac{d^4k \,
d^4l}{(2\pi)^{8}}\,\left[ 1+(-1)^n \frac{k^{2n}}{\Lambda^{2n}}
\right]^{-1} \left[ 1+(-1)^n \frac{l^{2n}}{\Lambda^{2n}}
\right]^{-1}\, \times\nonumber\\
&&\quad \quad \times\frac{1}{k^2\, l^2\, (k+q)^2\, (q+l)^2
(k+q+l)^2} \times \nonumber\\
&& \qquad \times \Bigg( -2 + \frac{(2q+k+l)_{\mu}
(k+q)^{\mu}}{(k+q)^2} + \frac{(2q+k+l)_{\mu} (l+q)^{\mu}}{(l+q)^2}
\Bigg);\\
\label{f-3} && f_3(q^2)= -4 i \, e^6 \int \frac{d^4 k\, d^4 l\,
d^4 r}{(2\pi)^{12}}\,\left[ 1+(-1)^n \frac{k^{2n}}{\Lambda^{2n}}
\right]^{-1} \left[ 1+(-1)^n \frac{l^{2n}}{\Lambda^{2n}}
\right]^{-1} \times\nonumber\\
&& \times \left[ 1+(-1)^n \frac{r^{2n}}{\Lambda^{2n}}
\right]^{-1}\, \frac{1}{k^2\, l^2\, r^2\, (k+q)^2\, (q+l)^2\,
(q+k+l)^2}\,
\Bigg(\frac{1}{(k+q+r)^2} + \nonumber \\
&& + \frac{1}{2(k+l+q+r)^2} - \frac{(2q+k+l)^{\mu}
(k+q)^{\mu}}{2(k+q)^2\, (k+q+r)^2} - \frac{(2q+k+l)^{\mu}
(l+q)^{\mu}}{2(l+q)^2\, (k+q+r)^2} - \vphantom{\Bigg(} \nonumber \\
&&  - \frac{(2q+k+l)^{\mu} (k+q)^{\mu}}{2(k+q)^2\, (k+l+q+r)^2}\,
- \frac{(2q+k+l)^{\mu} (k+q+r)^{\mu}}{2(k+q+r)^4} \Bigg).
\end{eqnarray}

\begin{eqnarray}\label{F_1}
&& F_1 (q^2) = -\, \frac{ie^2}{8} \int \frac{d^4 k}{(2\pi)^4}\,
\left[ 1+(-1)^n \frac{k^{2n}}{\Lambda^{2n}}
\right]^{-1}\frac{1}{k^2\, (k+q)^4};\\
\label{F_2} && F_2(q^2) = \frac{1}{4}\, e^4\int \frac{d^4k \,
d^4l}{(2\pi)^{8}}\,\left[ 1+(-1)^n \frac{k^{2n}}{\Lambda^{2n}}
\right]^{-1} \left[ 1+(-1)^n \frac{l^{2n}}{\Lambda^{2n}}
\right]^{-1}\times\nonumber\\
&&\times \frac{1}{k^2\, l^2\, (k+q)^2\, (q+l)^2 (k+q+l)^2} \Bigg(
4 -2\, \frac{(q+l)^2}{(k+q)^2} -\, \frac{(2q+k+l)^2}{(q+k+l)^2}
\Bigg).
\end{eqnarray}

\noindent (An expression for $F_3$ is not presented, because it is
too large.)


\end{document}